\documentclass[prl,groupedaddress,twocolumn,superscriptaddress]{revtex4}

\def\a1{$a_{1}\:$}
\def\ax2{$a_{2}\:$}

\def\A{\AA$^{-1}\:$}
\def\Am3{\AA$^{-3}\:$}
\def\tc{$T_c \;$}

\def\sqw{$S(Q,\omega)\:$}

\def\he4{$^4$He}
\def\he3{$^3$He}
\def\rs{$\rho_S\:$}

\def\rst{$\rho_S(T)\:$}

\def\no{$n_{0}\:$}

\def\nk{$n({\bf k})\:$}
\def\nstar{$n^{*}({\bf k})\:$}

\def\Ke3{$\langle{K_3}\rangle\:$}
\def\h2o{H$_2$O}
\def\d2o{D$_2$O}

\def\jqy{$J(Q,y)\:$}

\def\jia{$J_{IA}(y)\:$}
\def\rqy{$R(Q,y)\: $}

\def\ccmol{cm$^3$/mol\hspace{.1cm}}
\def\alt{$\overline{\alpha}_2\:$}
\def\alf{$\overline{\alpha}_4 \:$}
\def\ath{$\overline{a}_{3}/\lambda\:$}

\def\del{$\frac{\overline{\alpha}_4}{{\overline{\alpha}_2}^2}\:$}

\def\mutb{$\overline{\mu}_3 \:$}

\usepackage{graphicx}

\begin{document}

\title{Bose-Einstein Condensation in solid $^4$He}

\author{S.O. Diallo}
\affiliation{Department of Physics and Astronomy, University of Delaware, Newark, Delaware 19716-2593, USA}
\author{J.V. Pearce}
\affiliation{National Physical Laboratory, Teddington, United Kingdom}

\author{R.T. Azuah}
\affiliation{NIST Center for Neutron Research, Gaithersburg, Maryland 20899-8562, USA}
\affiliation{Department of Materials Science and Engineering, University of Maryland, College Park, Maryland 20742-2115, USA}
\author{O. Kirichek}
\affiliation{ISIS Spallation Neutron Source, Rutherford Appleton Laboratory, Chilton, Didcot, OX11 0QX, United Kingdom}
\author{J.W. Taylor}
\affiliation{ISIS Spallation Neutron Source, Rutherford Appleton Laboratory, Chilton, Didcot, OX11 0QX, United Kingdom}
\author{H.R. Glyde}
\affiliation{Department of Physics and Astronomy, University of Delaware, Newark, Delaware 19716-2593, USA}
\date{\today}
\begin{abstract}
We present neutron scattering measurements of the atomic momentum distribution, $n({\bf k})$, in solid helium under a pressure $p =$ 41 bars and at temperatures between 80 mK and 500 mK. The aim is to determine whether there is Bose-Einstein condensation (BEC) below the critical temperature, $T_c$ = 200 mK where  a superfluid density has been observed. Assuming BEC appears as a macroscopic occupation of the $k =$ 0 state below $T_c$, we find a condensate fraction of $n_0$ = (-0.10 $\pm$1.20)\%  at $T =$ 80 mK and $n_0$ = (0.08$\pm$0.78)\% at $T=$120 mK, consistent with zero. The shape of $n({\bf k})$ also does not change on crossing $T_c$ within measurement precision. 
  \end{abstract}
\maketitle
In 2004, Kim and Chan \cite{Kim:04b,Kim:04a} reported the spectacular observation of a superfluid density in solid helium below a critical temperature, \tc. The superfluid density, \rs, is observed as a non-classical rotational inertia (NCRI) in a torsional oscillator (TO) containing solid helium. The percent of solid that has a NCRI and is decoupled from the rest of the solid was \rst $\simeq$ 1.5 \% at temperature $T=$ 50 mK in commercial grade purity $^4$He which contains typically 0.3 ppm of $^3$He, where \tc = 200 mK. All other impurities are frozen out. A \rs was observed in both bulk solid helium \cite{Kim:04b} and in solid confined in porous media (Vycor) \cite{Kim:04a}. The magnitude of \rs varies somewhat from solid sample to solid sample \cite{Kim:06}. A superfluid density was observed in solids at pressures between $p\sim$ 25 bars near the melting line and 135 bars with \rs taking its maximum value of 1.5 \% at $p \sim$ 50 bars.

	This remarkable result has been confirmed in independent TO measurements \cite{Shirahama:06, Rittner:06,Choe:06}. Rittner and Reppy \cite{Rittner:06} find that \rs can be significantly reduced by annealing the solids near their melting temperatures with \rs reduced to zero in some cases. Similarly, Shirahama and coworkers \cite{Shirahama:06} report a reduction in \rs of up to 50 \% by annealing. Macroscopic superflow was not observed in helium in Vycor \cite{Day:05} and bulk helium \cite{Day:06}. However, Sasaki {\it et al.} \cite{Sasaki:06} have observed macroscopic superflow in those solids which contain grain boundaries that extend across the solid. This unexpected result suggests that there is indeed superflow and that it is along or associated with grain boundaries. Superflow related to grain boundaries \cite{Sasaki:06}, the variation of \rs from sample to sample \cite{Kim:06} and the reduction of \rs by annealing \cite{Rittner:06,Shirahama:06} suggest that a superfluid density may be associated with macroscopic defects that extend across or whose impact extends across the whole solid.

 In liquid helium, Bose-Einstein condensation (BEC) and superfluidity are observed together. Indeed superflow can be shown to follow from BEC \cite{Baym:69, Nozieres:book}. It can also be shown to arise from long range atomic exchanges \cite{Ceperley:86}. In the liquid, both in bulk \cite{Glyde:00} and in Vycor \cite{Azuah:03}, BEC is observed as a macroscopic occupation of the $k=0$ state in \nk, as expected for a translationally invariant system. Solid helium is also translationally invariant, as is a gas of vacancies in the solid. Thus, for certain models of superflow we anticipate that BEC will appear as a macroscopic occupation of $k = 0$ in \nk. In this context we look for an enhancement of \nk at $k\sim$ 0 below \tc. We also look for a change in shape of \nk below \tc.  The central result is that we observe no increase in \nk at $k\sim0$ nor any change in shape of \nk as the temperature is lowered below \tc.

In 1969, Andreev and Lifshitz \cite{Andreev:69} proposed that helium could be a supersolid if the solid contained vacant sites in the ground state. Essentially, the ground state, zero point vacancies could form a Bose gas at $T\sim$ 0 K in which the Bose-Einstein condensate fraction, \no, and \rs are both approximately 100 \%. 
While thermally activated vacancies have been observed \cite{Simmons:94}, ground state vacancies have not. Chester \cite{Chester:70} proposed that superflow in solid helium may be possible because the solid is well described by fluid like wave functions that support superflow. Leggett \cite{Leggett:70} examined superflow via long range exchanges of atoms within a perfect helium solid and found superflow possible but that \rs would be small, \rs $\sim$ 0.01 \%.

	Recent accurate path integral Monte Carlo calculations \cite{Ceperley:94} find that \rs  arising from long range atomic exchanges in a perfect crystal will be unobservably small. Similarly, BEC in bulk perfect crystals is predicted to be very small \cite{Boninsegni:06a,Clark:06}. Zero point vacancies in the ground state are predicted to be unstable \cite{Boninsegni:06a}. Essentially, vacancies migrate to a surface or coalese to create a surface and leave the crystal. However, if solid helium is held in an amorphous rather the equilibrium crystal state, both \rs and \no take significant values \cite{Boninsegni:06a}. With these predictions, it is interesting to search for BEC where \rs is observed.

The solid $^4$He investigated in this experiment was grown using the blocked capillary method. Commercial grade purity $^4$He (0.3 ppm $^3$He) was introduced into a cylindrical Al sample cell of 20 {mm}
diameter and 62 {mm} height at a temperature $T\simeq$ 3 K to a pressure of $p\simeq$ 70 bars. At constant $p$, the temperature was reduced using an Oxford Instruments Kelvinox VT dilution refridgerator until solid formed in the capillary and blocked the cell. The block was observed at filling $p=$ (69.8 $\pm$0.2) bars and $T=$ (2.79 $\pm$0.02) K, which corresponds to liquid on the melting line at a molar volume $V_m$ = (20.01 $\pm$0.02) \ccmol \cite{Driessen:86}.   
The blocked cell was further cooled and neutron inelastic scattering measurements at high momentum transfer were made in the solid hcp phase at 80 mK, 120 mK, 300 mK and 500 mK on the MARI spectrometer at the ISIS neutron facility.
\begin{figure}
\includegraphics[width=0.85\linewidth]{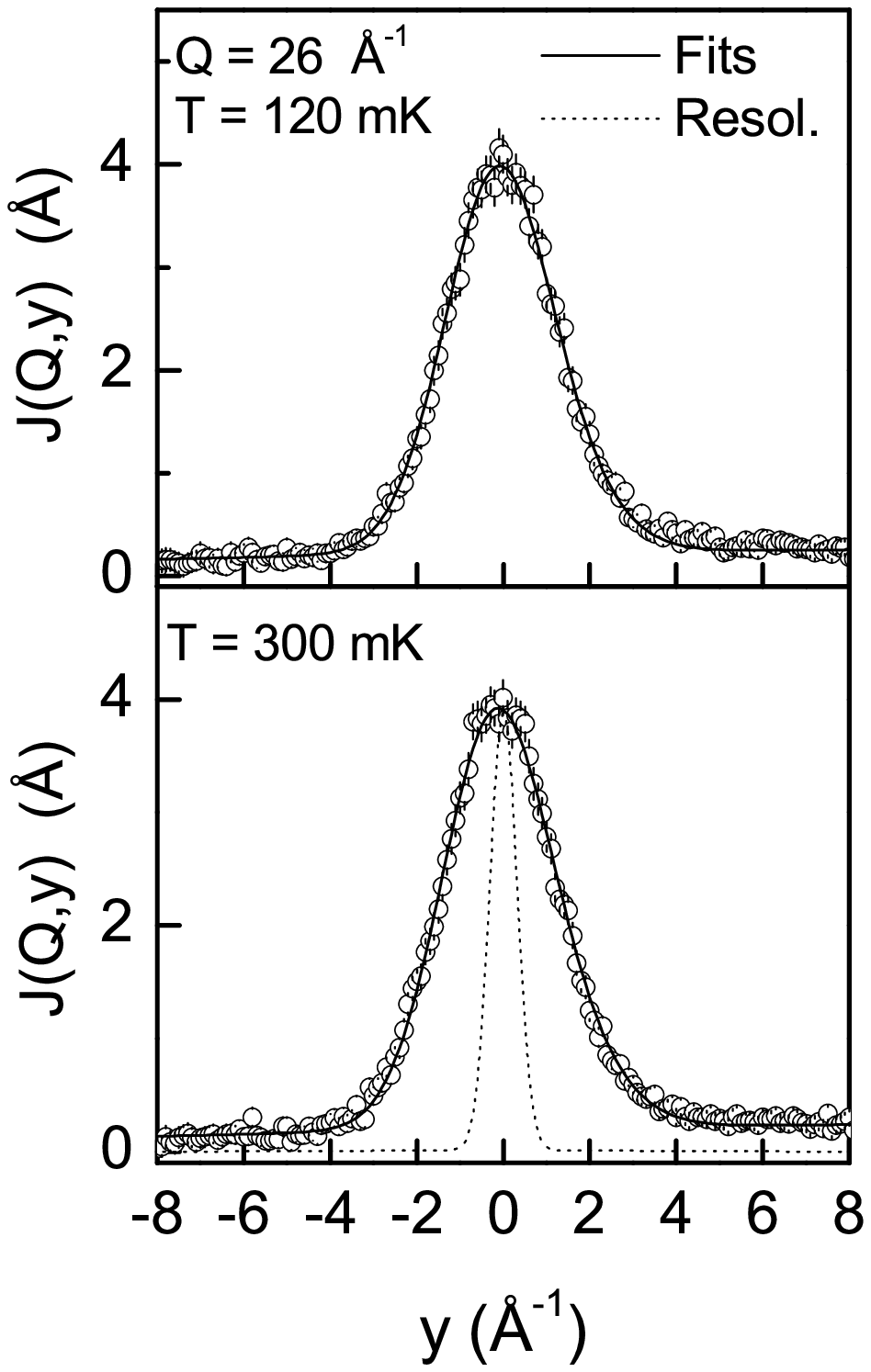}
\caption{\footnotesize 
Observed $J(Q,y)$  of solid $^4$He at V = 20.01 \ccmol (open circles) folded with the instrument resolution at momentum transfer $Q=$ 26 \A  and temperatures $T=120$ (top) and 300 mK (bottom). The solid lines are fits of the Convolution Approach (CA) with Final State (FS) function taken from liquid $^4$He (Ref. \cite{Glyde:00}), shape of \nk from Ref. \cite{Diallo:04} and width \alt parameter as the single free fitting parameter.  The data peaks below $y = 0$ because of FS effects. The dotted line is the MARI instrument resolution function.}
\label{fig_b}
\end{figure}
\begin{figure}
\includegraphics[width=0.95\linewidth]{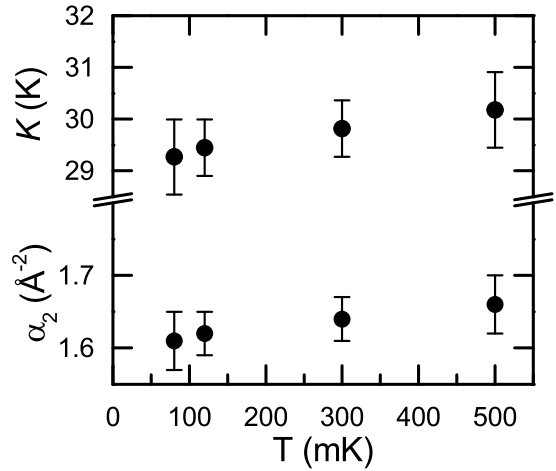}
\caption{
\footnotesize  The mean square atomic momentum, \alt = $\langle k_Q^2\rangle$ along an axis and the atomic kinetic energy $K = (3\lambda/2)$\alt where $\lambda$ = 12.12 K \AA$^2$ of solid helium at V = 20.01 {cm}$^3$/mole versus temperature.
 } 
\label{fig_c}
\end{figure}

Single atom dynamics is observed in the dynamic struture factor \sqw at high momentum transfer, $\hbar Q$. Specifically, at $\hbar Q \rightarrow \infty$, where the Impulse Approximation (IA) is valid, the energy transfer $\hbar \omega$ in \sqw is Doppler broadened by the atomic momentum distribution \nk. In this limit, it is convenient to express $\omega$ in terms of the \lq {\it y} scaling' wave vector variable, 
$y=(\omega-\omega_R)/v_R$ where $\omega_R=\hbar Q^2/2m$ and $v_R=\hbar Q/m$, and to present the neutron inelastic data as $J(Q,y)=v_R S(Q,\omega)$. Including final state (FS) effects which are small but not negligible at the $Q$ values investigated here,

\begin{equation}
J(Q,y)=\int dy'R(Q,y-y')J_{IA}(y')
\end{equation}   
\noindent where $R(Q,y)$ is the FS broadening function and 
\begin{equation}
J_{IA}(y)=\int d{\bf k} n({\bf k}) \delta(k_Q-y)= n_Q(y)
\end{equation}
\noindent is the IA to $J(Q,y)$. Specifically, \jia is \nk projected along ${\bf Q}$ denoted the longitudinal momentum distribution \cite{Silver:89}.

	Fig. \ref{fig_b} shows the observed \jqy at wavevectors $Q =$ 26 \A and temperatures $T =$ 120 mK and $T=$ 300 mK. The observed \jqy includes the MARI instrument resolution function which is shown separately as a dotted line in Fig. \ref{fig_b}. The solid line is a fit of a model to the data as described below. 
We were able to determine at most two free parameters in model fits to the data. 

To obtain a condensate fraction, we assumed a model \nk of the form,
\begin{equation}
	n({\bf k}) = n_0 \delta ({\bf k})  +  (1-n_0) n^{*}({\bf k}), 
\label{model}
	         \end{equation}
where \nstar is the momentum distribution of the atoms above the condensate in the $k > 0$ states. To proceed, we assume (1) that the shape of \nk is the same as observed previously in solid helium at a somewhat lower pressure \cite{Diallo:04} and (2) that the FS function \rqy in Eq.(1) is the same as observed \cite{Glyde:00} in liquid helium. The free parameters (two) in the model are then \no and the width, $\overline\alpha_2$, of the Gaussian component of \nstar. A condensate component appears as an additional intensity unbroadened by \nstar in \jia at $y=0$.

\begin{figure}
\includegraphics[width=0.75\linewidth]{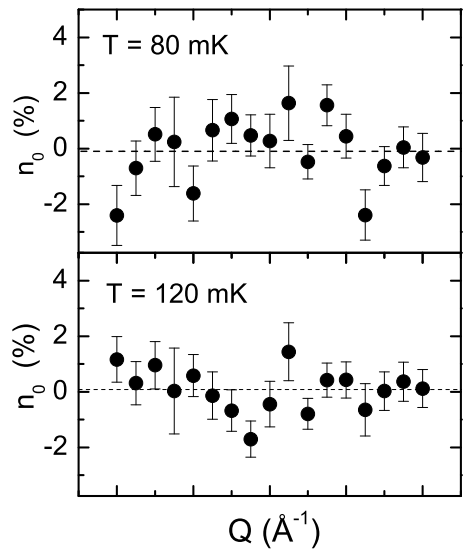}
\caption{\footnotesize Condensate fractions obtained by fitting the Convolution Approach (CA) at $T=$ 80 mK and 120 mK with the width of \nk (\alt) held fixed at its values shown in Fig.\ref{fig_c}.  We find \no = (-0.01 $\pm$1.20)\% at $T=$ 80 mK and \no = (0.08$\pm$0.78)\%.
}
\label{fig_d}
\end{figure} 
	We first fit the model \nk to the data at 500, 300, 120 and 80 mK assuming \no = 0 at all temperatures. The \no is expected to be zero at 300 and 500 mK. The resulting values of $\overline\alpha_2$ and the associated atomic kinetic energy ($K = (3 \hbar^2/2m) \overline\alpha_2$) are shown in Fig. \ref{fig_c}. The resulting $\overline\alpha_2$ and $K$ decrease somewhat with decreasing temperature. In a recent measurement, Adams {\it et al.} \cite{Adams:07} find $K$ independent of temperature within precision. 

If superflow is associated with defects in the solid such as vacancies, we anticipate that BEC is similarly associated with these defects, perhaps in a vacancy gas. In this event, the atomic kinetic energy, $K$, of the majority of the atoms may be largely unaffected by the BEC in the defects. To obtain \no at 80 mK and 120 mK within this picture we kept $\overline\alpha_2$ fixed at the values obtained above and refitted model (\ref{model}) with \no as a free parameter. The fitted values of \no are shown in Fig. \ref{fig_d}. The variation of \no with Q reflects the statistical error in \no. The mean values are \no = (-0.1 $\pm$ 1.2) \% at 80 mK and  \no = (0.08 $\pm$ 0.78) \% at 120 mK.

If, in contrast, the superflow and BEC lie within the bulk of the solid, we anticipate, as in liquid $^4$He, that the atomic kinetic energy, K, will decrease below \tc as a result of BEC.  The observed decrease in $K$ below \tc  has been used to estimate \no  in liquid $^4$He \cite{Sears:83, Mayers:97b}. To address this case, we kept $\overline\alpha_2$ (kinetic energy arising from $n^{*}({\bf k})$, constant at the value obtained at 300 mK and refitted model (\ref{model}) at 80 mK and 120 mK to obtain \no. The resulting values are \no = (0.8 $\pm$ 1.2) \% at 80 mK and \no = (0.76 $\pm$ 0.77) \% at 120 mK. This method assumes that all the drop in K below \tc = 200 mK arises from BEC. To normalize these \no values for changes in $K$ from other sources, we also determined \no at 300 mK. That is, we kept the $\overline\alpha_2$ fixed at its 500 mK value and refitted the model to determine \no at 300 mK giving \no = (0.63 $\pm$ 0.77) \%. Since \no at 300 mK is zero, we expect the \no values below \tc to be overestimated by approximately 0.6 \%.  Normalizing for this overestimate, we arrive at \no $\simeq (0.2 \pm 1.2)$ \% at 80 mK and \no $\simeq  (0.1 \pm 0.8)$ \% at 120 mK. If we try to determine both $\overline\alpha_2$ and \no simultaneously, we get essentially the same  values as before, with only larger error bars;     e.g. \no= (0.74 $\pm$ 1.01) \% at $T=$ 120 mK. Thus all methods give similar values of \no.
                                                  
\begin{figure}
\includegraphics[width=0.85\linewidth]{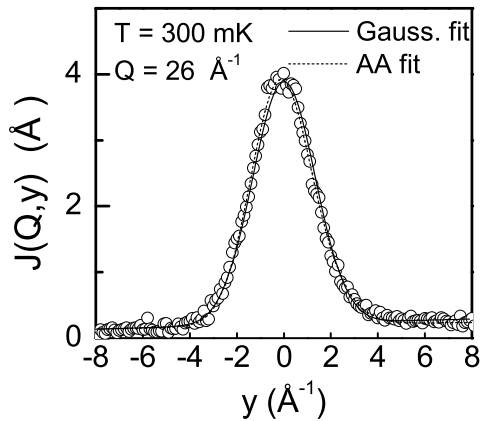}
\caption{\footnotesize  Fits of the Additive Approach (AA ) which includes FS effects and a simple Gaussian to the observed \jqy.
}
\label{fig_e}
\end{figure} 
\begin{figure}
\includegraphics[width=0.8\linewidth]{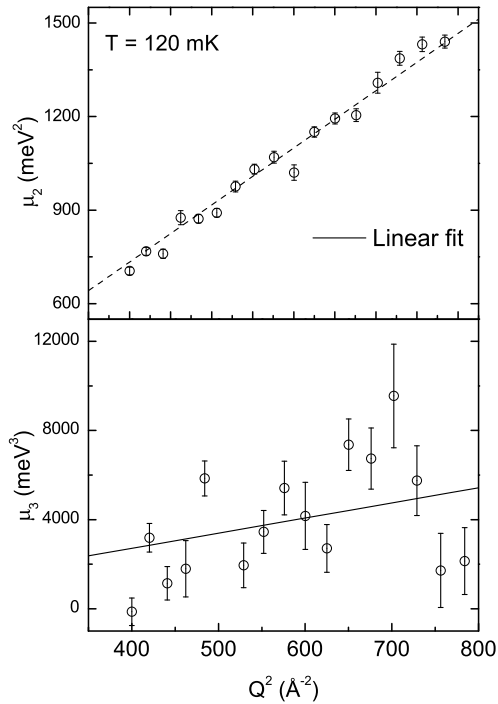}
\caption{ \footnotesize Fitting parameters   $\hbar^2\mu_2  =  ({\lambda}Q)^2 \ \bar{\alpha}_2$ and $\hbar^3\mu_3  =  ({\lambda}Q)^2 \ \bar{a}_3$ in the AA giving \alt = 1.67 {\AA$^{-2}$} and \ath = 6.21 {\AA}$^{-4}$. } 
\label{fig_f}
\end{figure}

To investigate a possible change in shape of \jqy or \nk on crossing \tc = 200 mK, we fit the additive approach (AA) \cite{Andersen:97} to the data. In this model
 to lowest order, $J(Q,y)$ is \cite{Azuah:03,Andersen:97}
\begin{eqnarray}
J(Q,y)&=&\Big[1- \frac{\overline\mu_3}{2 \overline\alpha_2 ^{3/2}}\Big(x-\frac{x^3}{3}\Big)+\frac{\overline\mu_4}{8\overline\alpha_2^2}\Big(1-2x^2\nonumber\\
       &&+\frac{x^4}{3}\Big)\Big]J_G(x)
\end{eqnarray}
where $J_G=\frac{1}{\sqrt{2\pi \overline\alpha_2}}\exp (\frac{-x^2}{2})$ and $x=y/{\overline{\alpha}_2}^{1/2}$. The second term in (4)is the leading FS term and the third term is the leading correction to a Gaussian \nk plus a FS term. The fitting parameters are $\overline\alpha_2$, $\overline\mu_3$ = $\overline{a}_3/\lambda Q$, and $\overline\mu_4$  =  $\overline{\alpha}_4 + \overline{a}_4/(\lambda Q)^2$. Previously we found $\overline{a}_4$ = 0 \cite{Glyde:00,Azuah:03,Silver:89}. The three remaining parameters are \alt, \mutb, and \alf. The kurtosis of \nk is $\delta$ = \del. 

Fig. \ref{fig_e} compares AA and simple Gaussian fits to data. The AA gives a better fit with the peak position lying at $y<0$ in agreement with the data. We were able to determine only two parameters, e.g. $\hbar^2\mu_2  =  ({\lambda}Q)^2 \overline{\alpha}_2$ and $\hbar^3\mu_3  =  ({\lambda}Q)^2 \ \overline{a}_3$  where  
$\lambda = \hbar^2/m = 1.0443$ meV \AA$^2$ = 12.12 K \AA$^2$. These parameters are plotted in Fig. \ref{fig_f} and show the expected $Q$ dependence. We found again a smooth change in \alt with temperature as shown in Fig.\ref{fig_c} and found \ath independent of temperature.  We tried keeping \alt fixed and fitting for \mutb and \alf but we were not able to determine \alf.  $\delta$ = 0 or $\delta =0.4 $ fit equally well. Thus we find no change in the shape of \jqy or \nk at \tc within precision. 

The condensate fraction, $n_0$, of a perfect crystal is expected to be very small \cite{Clark:06, Boninsegni:06a}, \no $< 10^{-8}$. In a perfect crystal, BEC requires double occupation of a lattice site which has very high energy and is therefore very improbable. 
If there are vacant sites, \no in the crystal is dramatically increased, to \no $\simeq$ 0.23 \% for a vacancy concentration $c_V \simeq 1$ \% \cite{Galli:06}. If the solid is frozen in a non-equilibrium amorphous state, \no $\simeq$ 0.5 \% \cite{Boninsegni:06b}. The \no in the amorphous state is relatively insensitive to density although \rs decreases significantly with increasing density \cite{Boninsegni:06b}.

If the vacancies are treated as an ideal Bose gas in which \no $\simeq$ 100\% and \rs$\simeq$ 100\%, a vacancy concentration gas 
 $c_V \sim$ 1.5 \% is needed to reach \rs = 1.5 \% in the solid. This also predicts \no $\sim$ 1.5 \%. This large value of \no and this simple model appears to be excluded by our observed values, of \no = (-0.10 $\pm$1.20)\%. In contrast, the \no values within bulk solid helium including vacancies \cite{Galli:06} or in extended amorphous regions \cite{Boninsegni:06b}, whether in equilibrium or not, are consistent with our observed value.

	In summary, we have determined the BEC condensate fraction in commercial grade purity solid helium at pressure 41 bars and molar volume 20.01 {cm}$^3$/mole using inelastic neutron scattering. We find a condensate fraction \no = ( -0.10 $\pm$1.20)\%  below \tc = 200 mK consistent with zero. We also find no change in the shape of the atomic momentum distribution on crossing \tc.
\bibliographystyle{unsrt}

\end{document}